\documentclass[twocolumn]{aastex631}

\usepackage{multirow}
\usepackage{rotating}
\usepackage{amsmath}

\makeatletter
\newcommand{\superimpose}[2]{{%
  \ooalign{%
    \hfil$\m@th#1\@firstoftwo#2$\hfil\cr
    \hfil$\m@th#1\@secondoftwo#2$\hfil\cr
  }%
}}
\makeatother
\newcommand{\notr}{\mathrel{\mathpalette\superimpose{{\mathrm{r}}{\backslash}}}}
\newcommand{\nots}{\mathrel{\mathpalette\superimpose{{\mathrm{s}}{\backslash}}}}

\begin{document}

\title{Classifying metal-poor stars with machine learning using nucleosynthesis calculations}

\author[0000-0002-3305-4326]{Nicole Vassh}
\affiliation{TRIUMF, 4004 Wesbrook Mall, Vancouver, BC V6T 2A3, Canada}

\author[0009-0001-3471-0364]{Yilin Wang}
\affiliation{University of British Columbia, Vancouver, BC V6T 1Z1, Canada}
\affiliation{TRIUMF, 4004 Wesbrook Mall, Vancouver, BC V6T 2A3, Canada}

\author[0000-0002-5126-7878]{Richard M. Woloshyn}
\affiliation{TRIUMF, 4004 Wesbrook Mall, Vancouver, BC V6T 2A3, Canada}

\author[0000-0002-9735-6147]{Michelle P. Kuchera}
\affiliation{Davidson College, Davidson, North Carolina 28035}

\author{Maude Larivi\`ere}
\affiliation{University of British Columbia, Vancouver, BC V6T 1Z1, Canada}
\affiliation{TRIUMF, 4004 Wesbrook Mall, Vancouver, BC V6T 2A3, Canada}

\author{Kayle Majic}
\affiliation{University of Victoria, Victoria, BC V8W 2Y2, Canada}

\author[0000-0002-9986-8816]{Benoit C\^ot\'e}
\affiliation{University of Victoria, Victoria, BC V8W 2Y2, Canada}
\affiliation{Konkoly Observatory, Konkoly Thege M. ut 15-17, Budapest 1121, Hungary}

%% Note that the \and command from previous versions of AASTeX is now
%% depreciated in this version as it is no longer necessary. AASTeX 
%% automatically takes care of all commas and "and"s between authors names.

%% AASTeX 6.31 has the new \collaboration and \nocollaboration commands to
%% provide the collaboration status of a group of authors. These commands 
%% can be used either before or after the list of corresponding authors. The
%% argument for \collaboration is the collaboration identifier. Authors are
%% encouraged to surround collaboration identifiers with ()s. The 
%% \nocollaboration command takes no argument and exists to indicate that
%% the nearby authors are not part of surrounding collaborations.

%% Mark off the abstract in the ``abstract'' environment. 

\begin{abstract}
We apply the capabilities of machine learning (ML) to discern patterns in order to classify metal-poor stars. To do so, we train an ML model on a bank of nucleosynthesis calculations derived from hydrodynamic simulations for events such as neutron star mergers where the rapid ($r$) neutron capture process can take place. Likewise we consider a bank of calculations from simulations of the slow ($s$) neutron capture process and also consider a few calculations for the intermediate ($i$) neutron capture process. We demonstrate that the ML does well overall in recognizing the $s$ process from the $r$ process, and after training on theoretical calculations ML stellar assignments match conventional labels 87\% of the time. We highlight that this method then points to stars that could benefit from additional observational measurements. We also demonstrate that the ML assigns some of the presently considered $i$-process stars to instead be of $r$ or $s$ in origin, but likewise, finds stars currently labeled as $s$ to be potentially more aligned with $i$ enrichment. This first application of ML to classify metal-poor star enrichment using theoretical nucleosynthesis calculations thus reveals the promise, and some challenges, associated with this new data-driven path forward.
%250 word limit for the abstract

\end{abstract}

%% Keywords should appear after the \end{abstract} command. 
%% The AAS Journals now uses Unified Astronomy Thesaurus concepts:
%% https://astrothesaurus.org
%% You will be asked to selected these concepts during the submission process
%% but this old "keyword" functionality is maintained in case authors want
%% to include these concepts in their preprints.
%%\keywords{Classical Novae (251) --- Ultraviolet astronomy(1736) --- History of astronomy(1868) --- Interdisciplinary astronomy(804)}

%% From the front matter, we move on to the body of the paper.
%% Sections are demarcated by \section and \subsection, respectively.
%% Observe the use of the LaTeX \label
%% command after the \subsection to give a symbolic KEY to the
%% subsection for cross-referencing in a \ref command.
%% You can use LaTeX's \ref and \label commands to keep track of
%% cross-references to sections, equations, tables, and figures.
%% That way, if you change the order of any elements, LaTeX will
%% automatically renumber them.
%%
%% We recommend that authors also use the natbib \citep
%% and \citet commands to identify citations.  The citations are
%% tied to the reference list via symbolic KEYs. The KEY corresponds
%% to the KEY in the \bibitem in the reference list below. 

\section{Introduction} \label{sec:intro}

The distinct nature of abundance patterns from different astrophysical processes has been highlighted for decades, ever since the foundational work of Burbidge, Burbidge, Fowler and Hoyle (B$^2$FH) \citep{B2FH} used patterns in the Solar abundance data to differentiate between the rapid ($r$) and slow ($s$) neutron capture processes. Nowadays, nucleosynthesis studies have more stars than the Sun to consider with stellar abundance data steadily growing. Metal-poor stars (i.e. [Fe/H]$\leq$-1) are especially interesting due to their abundances being connected to fewer astrophysical events, and very metal-poor stars (such as the [Fe/H]$\leq$-2 cases considered here) are even more appropriately compared to single events \citep{Frebel}. Such stars are typically assigned a classification based on the nucleosynthesis process presumed to have enriched it. Present criterion given by the JINAbase database \citep{JINAbase} is:

$\,$\newline
\noindent $r-I$: 0.3 $\leq$ [Eu/Fe] $\leq$ 1 and [Ba/Eu] $<$ 0, \newline
\noindent $r-II$: [Eu/Fe] $>$ 1 and [Ba/Eu] $<$ 0, \newline
\noindent $r-lim$: [Eu/Fe] $<$ 0.3, [Sr/Ba]$>$0.5, and [Sr/Eu]$>$0, \newline
\noindent $s$: [Ba/Fe] $>$ 1, [Ba/Eu] $>$ 0.5, and [Ba/Pb]$>$-1.5.
$\,$\newline

\noindent Such elemental ratio thresholds are inspired by key features of different processes as reported in the nucleosynthesis literature.   More recently, some stars previously labeled as ``r/s" (considered to not fit either $s$ or $r$ abundance patterns alone), have been re-classified as ``i" given the ability of calculations for an intermediate neutron capture process ($i$-process) to reproduce the observed stellar patterns \citep{Hampel2016,Ianiprocess,Paveliprocess}. However, such studies can overlook the plethora of $r$-process patterns possible in the diverse ejecta reported by hydrodynamic simulations.  

Although ML has been applied extensively in astrophysics, applications to star classification have been focused on analyzing large surveys of stars. For instance ML has been used to identify metal-poor candidates \citep{Hou2024,Matijevivc2017,Galarza2022,Hughes2022}, identify stars rich in certain elements \citep{Phosstars,Cstars}, analyze or estimate abundances / enrichment levels \citep{Hattori2025,denHartogh2023,Villagos2024,Hartwig2023}, and derive estimates of stellar parameters (e.g. metallicity, surface gravity, age) \citep{Gu2025,Khalatyan2024,Boulet2024,Hayden2022}. However, applications of ML to nucleosynthesis problems have been rather limited. Here studies have focused mainly on uses of emulators to enable inline nucleosynthesis within hydrodynamic simulations \citep{Grichener2024} and utilizing ML to predict nuclear properties of relevance to nucleosynthesis \citep{Mengke2024} and neutron star crusts \citep{UtamaPiekarewiczMass}. 

However, the possibility of using ML to point back to a particular source of enrichment for metal-poor stars has never before been explored, even though metal-poor stars are regularly used as a benchmark for nucleosynthesis calculations. In this work we explore the possibility that the data from simulations of astrophysical events and then post-processed to determine nucleosynthesis abundances can serve as a large training set with which an ML model can be trained to recognize whether a given elemental abundance pattern is consistent with an $r$ or $s$ process, thus laying the groundwork to use ML to classify the enrichment of stars. Additionally we demonstrate that this approach has promise in discerning whether some presently classified $i$-process stars may be more likely of $s$-process or $r$-process event origins, and vice versa.

To apply ML to determine stellar classifications from theoretical nucleosynthesis calculations, we consider several approaches. First, in Sec.~\ref{sec:nucleosets}, we describe our nucleosynthesis calculations that we take for training. In Sec.~\ref{sec:twoclass}, we discuss the use of a binary classifier to train on both $r$-process and $s$-process sets. Next, in Sec.~\ref{sec:oneclass}, we move towards a one-class classifier which is trained exclusively on one simulation data set ($r$ or $s$) and finds the areas of latent space in which results cluster themselves. After training, we then provide our ML models with the stellar abundances of several metal-poor stars and see which group our algorithm would consider each star to belong. We compare all our results to the current JINAbase classifications in Sec.\ref{sec:overallclassrs}. Lastly in Sec.\ref{sec:iprocess} we include some $i$-process calculations and presently labeled $i$-process stars and consider how our ML interprets their abundance patterns. We conclude in Sec.\ref{sec:conclusions}.

\section{Nucleosynthesis calculations as training sets}\label{sec:nucleosets}

\begin{figure}[t!]
\includegraphics[scale=0.38]{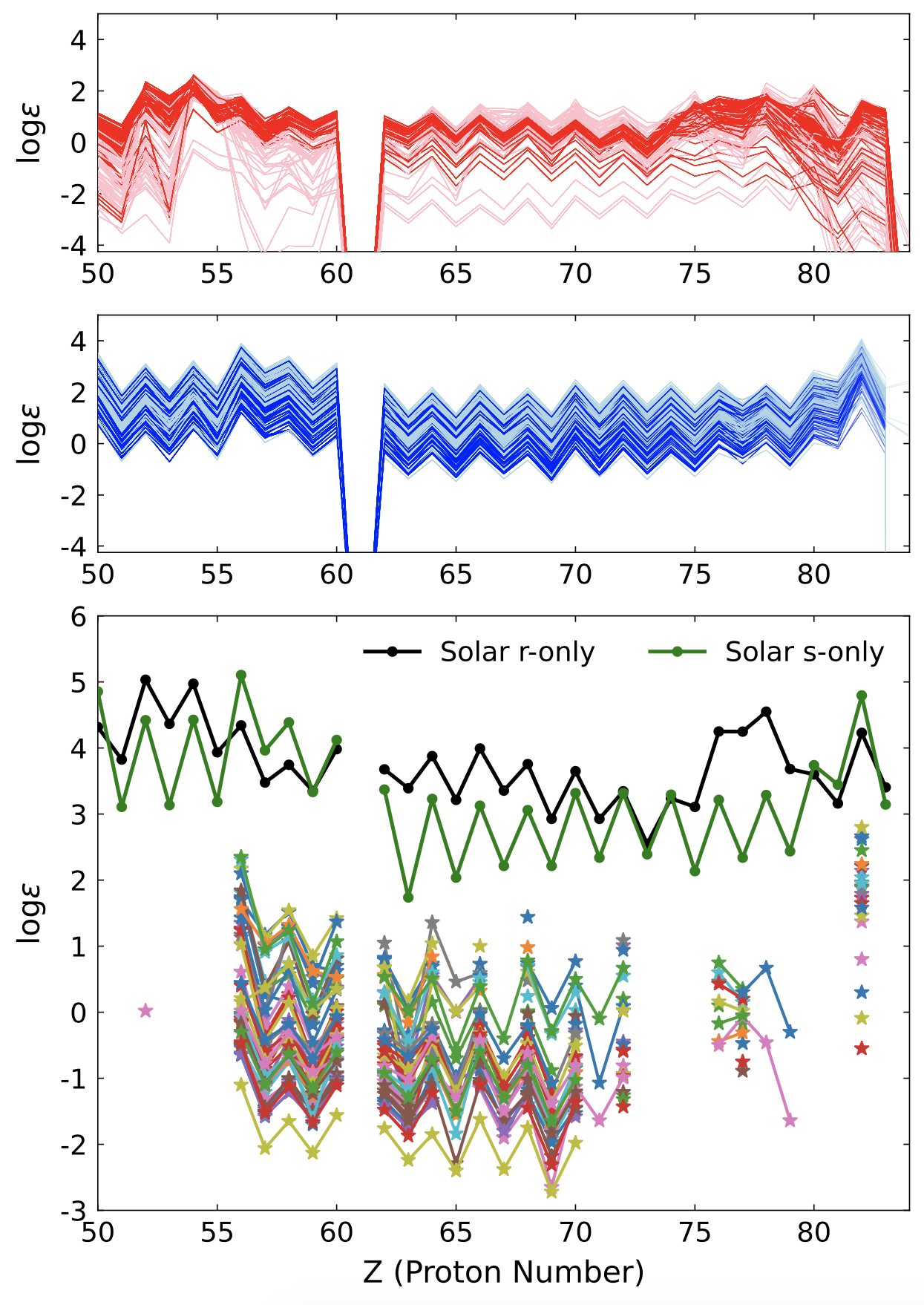}
\caption{(Top) Abundance patterns for the $r$-process calculations using the hydrodynamic simulations of Just et al. (pink) and Radice et al. (red). (Middle) Abundance patterns for the $s$-process calculations from Monash models (light blue) and FRUITY simulations (dark blue). (Bottom) Stellar abundance patterns for all metal-poor stars considered in this work (data from JINAbase) as compared to the Solar pattern broken down into its $r$ and $s$ contributions (from \cite{Sneden}).}
\label{fig:rsstars}
\end{figure}

Nucleosynthesis calculations which predict which elements are produced in different astrophysical scenarios can display a wide variety of abundance patterns (see Fig.~\ref{fig:rsstars}). For nucleosynthesis taking place at stellar sites, such as the $s$ process, variances largely stem from simulations considering different progenitor masses and metallicities. A wider diversity can be seen in $r$ process calculations by looking at nucleosynthesis results from the so-called ``tracers" from hydrodynamic simulations which collectively make up the picture of the total ejecta. We take these calculations as a bank of patterns for the elemental ratios that the overall ejecta could take on, and provide the ML model with all such cases rather than solely an averaged over mass-weighted total pattern. Since simulation predictions evolve, here we prefer an agnostic approach towards the overall distribution of $r$-process ejecta, and opt to maximize the diversity of our training data.

We gather several publicly available $s$-process nucleosynthesis predictions which also account for variation in multiple ways (two different simulation groups, different mass and metallicity grids for the AGB star, etc.). The set of 187 $s$-process patterns considered here (see Fig.~\ref{fig:rsstars}) combines both the Monash stellar models \citep{Monash1,Monash2,Monash3,Monash4,Monash5,Monash6} and the FRUITY simulations \citep{FRUITY}. For the $r$-process patterns, we consider all trajectories predicted to have non-zero mass ejection from hydrodynamic simulations of two distinct simulation groups that also represent distinct types of sites: Radice et al. \citep{Radice18} for merger dynamical ejecta and Just et al. \citep{Just} for accretion disk winds (we do not consider dynamical ejecta from Just et al.). Some differences are that the densities of Radice et al. tend to be higher and are typically more neutron-rich than Just et al. Note that Radice et al. considers 59 simulations by varying inputs such as equation of state, however available outflow data consists of mappings to parameterized trajectories. Thus these 59 simulations are captured by the same set of trajectories (all of which are considered here) mixed with different mass weightings when representing different outcomes\footnote{\url{https://zenodo.org/records/3588344}}. We then use these trajectories to perform the nucleosynthesis calculations with the reaction network PRISM \citep{BDFrp}, with nuclear physics inputs as in \cite{VasshFiss} (FRDM2012 model \citep{FRDM2012} with fission rates determined from FRLDM \citep{FRLDM} barriers along with AME2020 \citep{AME2020} and NUBASE2020 \citep{NUBASE2020} experimental data). 

For the $r$-process sets, we take only the cases that produce a main $r$ process between Z=54 to 83 and A=120 to 210. This reduces the set down to 420 in the case of the Radice et al. simulation and 64 for the Just et al. simulation. Since the available $s$-process data consists of 187 cases, in order to avoid any potential bias in our training that may be introduced by sets of unequal size, we take a subset of 188 $r$-process cases. To do this, we map the 420 Radice cases to a set of 124 by eliminating redundant cases using the STUMPY fast pattern matching algorithm\footnote{\url{https://stumpy.readthedocs.io/en/latest/Tutorial_Pattern_Matching.html}} which finds the nearest neighbor abundance pattern given the pattern of any particular tracer. We apply this to all tracers to find which have been uniquely selected as the neighbor, then reduce our data down to this set, and repeat the process until we obtain a set of the desired size (see Appendix for details). We then combine these 124 cases with the 64 from Just et al. to produce the set of 188 cases, all shown in Fig.~\ref{fig:rsstars}. After training on these nucleosynthesis calculations, we consider the metal-poor stars which are presently classified as $r$-, $s$-, or $i$- enriched (see Fig.~\ref{fig:rsstars}).

The first step in the ML classification is to decide on the feature set. To do so, we considered which elements in metal-poor stars tend to be reported in observational works and thus can be used for classification. This led to choosing 9 elements: Ba, La, Ce, Pr, Nd, Sm, Eu, Dy, Er (Z=56, 57, 58, 59, 60, 62, 63, 66, 68) as one feature set. Since lead (Pb, Z=82) abundances are often taken as a key discriminator between neutron capture processes, we explicitly investigate this by gathering a second feature set with Er replaced by Pb. This is an important test from a nucleosynthesis perspective, particularly given the wide range of predicted Pb abundances in $r$-process calculations. Note that the set of stars to be evaluated with our trained model is smaller in the Pb case due to fewer observations being available. Gathering stars with [Fe/H]$\leq$-2 yielded 115 cases and, to further blind our study, each star was assigned a number label; 43 of these stars reported abundances for the elements in our feature sets.

\section{Binary classifier}\label{sec:twoclass}

\begin{figure*}[t!]
\plotone{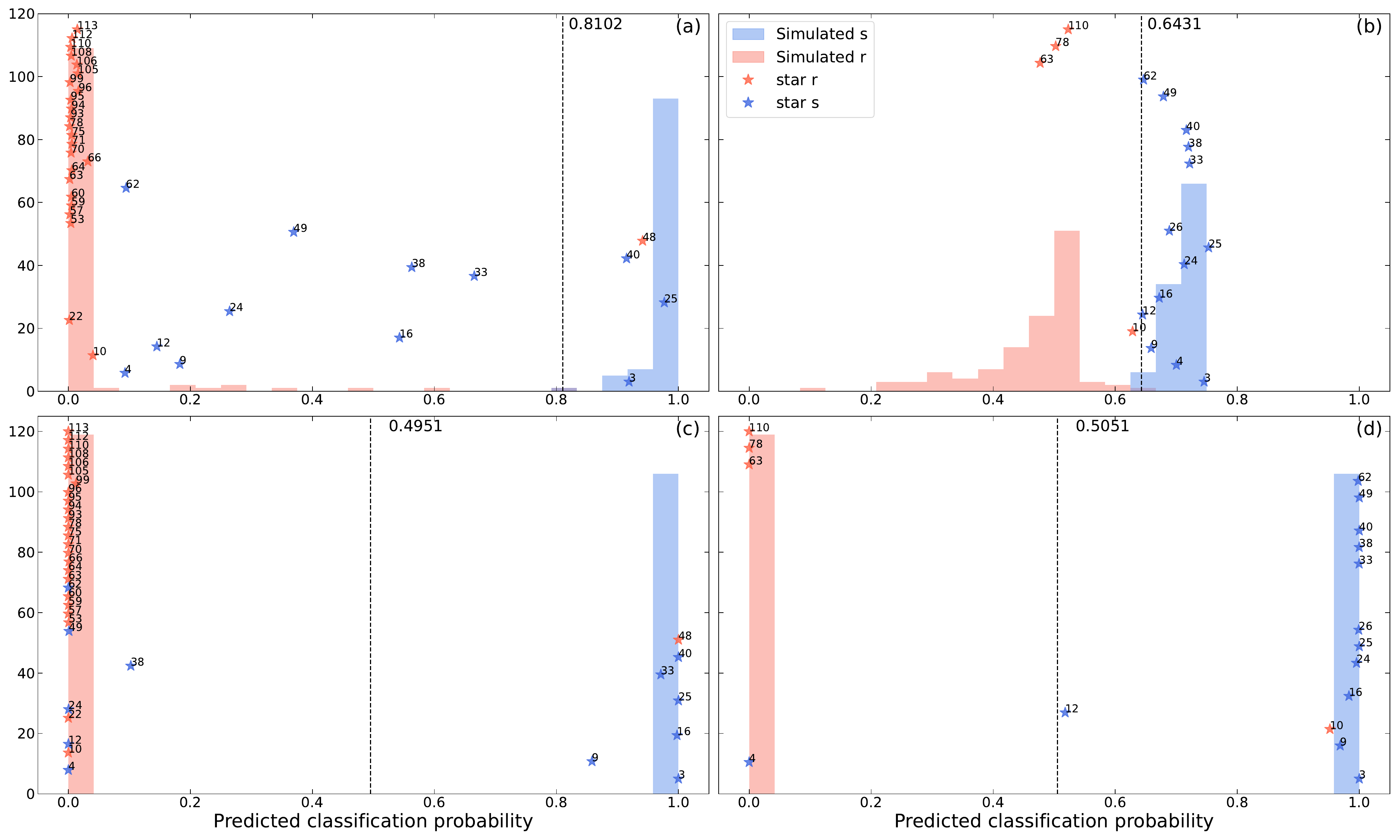}

\caption{Histograms showing the binary classification of the nucleosynthesis data (BC-L model here). Minimal separation for $r$ and $s$ histograms is shown for training including (a) Er and (b) Pb. Maximal separation is shown for training including (c) Er and (d) Pb. For results in lower panels, the model is trained for 10,000 epochs, and model weights from the epoch with the lowest validation loss are chosen as the final model. How the ML model classifies the stars in each case is also shown. For comparison, the color-coding of each star represents its JINAbase classification of which the ML model is never informed.
\label{fig:BC4panel}}
\end{figure*}

We first evaluate how a binary classifier (BC) performs in deciphering $r$-process versus $s$-process patterns. For this supervised ML model, the training, validation, and test data consisted of the sets of 187 $s$-process abundance patterns (labeled as 1) and 188 $r$-process patterns (labeled as 0) described in Sec.~\ref{sec:nucleosets}. We consider a fully connected shallow neural network (one input layer, one hidden layer, one output layer). The input layer contains 9 nodes that take in the normalized elemental abundances. The hidden layer contains 25 nodes in the case of the larger model (BC-L) and 5 nodes in the smaller model (BC-S). The output layer contains a single node which produces an output value onto which a Sigmoid activation function is applied, providing the final prediction between 0 and 1 (corresponding to $r$ and $s$, respectively). 

As shown in the top panel of Fig.~\ref{fig:BC4panel}, we first evaluate results at ``minimal separation", found by tracking the output values (predicted between 0 and 1) of cases in the training set, as well as their true labels (either 0 or 1). We find the last instance at which the largest predicted value for cases with a true label of 0 is larger than the smallest predicted value of cases with a true label of 1, and define the next epoch as the point of minimal separation. With further training activation function values are driven toward 0 ($r$-process) and 1 ($s$-process) (out to 10,000 epochs total was considered). Then the ``maximally separated" state (bottom panels of Fig.~\ref{fig:BC4panel}) refers to the overall best-trained state of the model, where the loss on the validation set is at its global minimum. We use a Binary Cross Entropy loss function to compute the error margin between predicted output and the true label for binary class predictions. We have also performed test variations of other hyper parameters such as the learning rate. Results from both models were fairly robust throughout hyper parameter tuning, with the BC consistently able to achieve near perfect separation within hundreds of epochs. 

After training we expose the model to the stellar data. We evaluate the activation values predicted for each star to see if the ML has grouped them into the $r$ or $s$ category. In Fig.~\ref{fig:BC4panel}, for comparison we also color-code each star based on its classification as defined by JINAbase. It is important to note that the ML is never informed of the database's classification. As can be seen in Fig.~\ref{fig:BC4panel}, when Er is used in the training set many stars are found near the $r$-process peak in the histogram, but when Pb is instead included, most stars are located near the $s$-process histogram peak. However, this result discovers a bias already present in the raw stellar data: most of the stars with abundances featuring Pb are those that have been classified as $s$-process or $i$-process stars. 

\begin{figure*}[!ht]
\centering
\includegraphics[scale=0.38]%{Figures/OCC_4panel_final1.pdf} 
{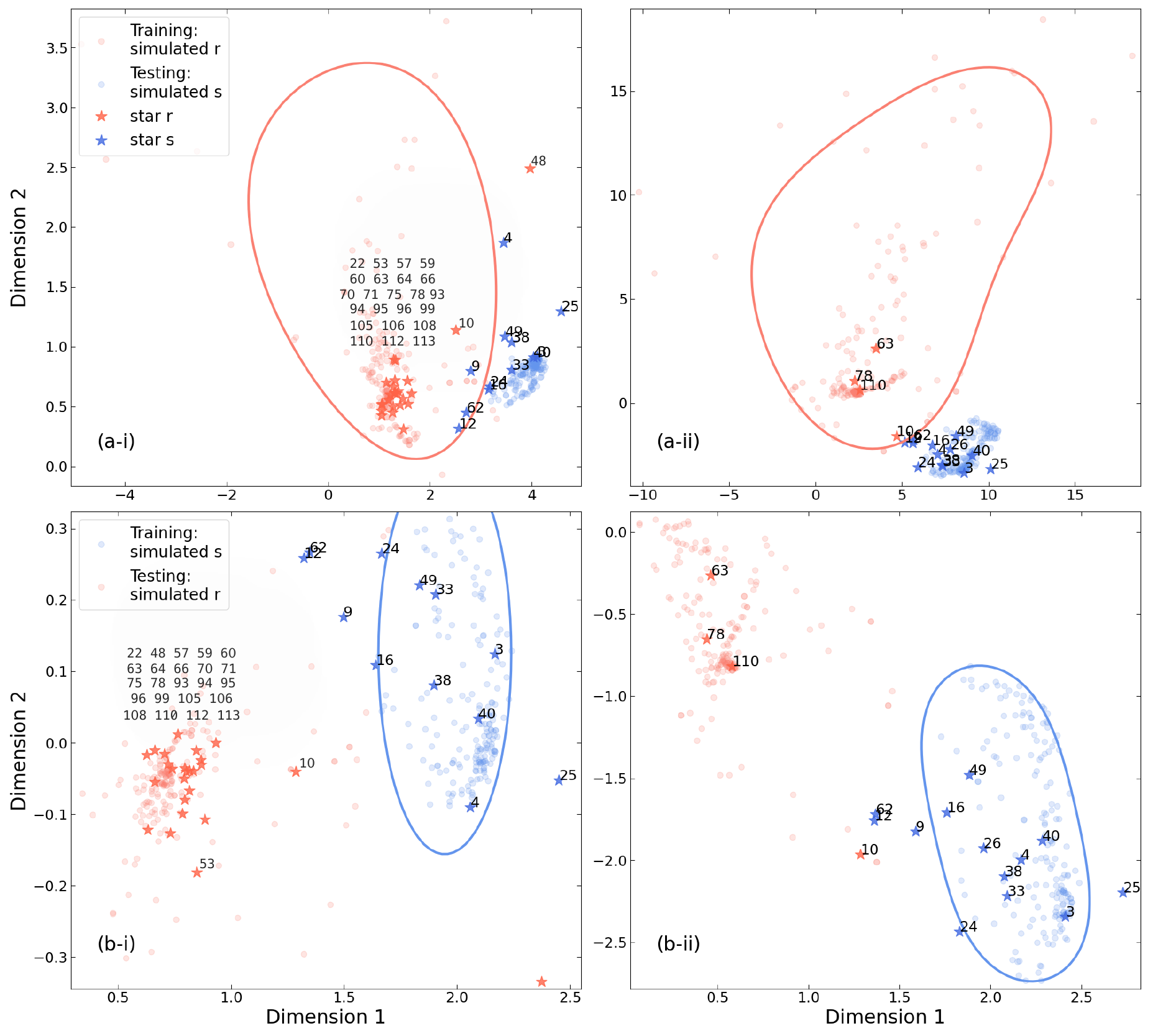} 
\caption{The two-dimensional latent space values (labeled as dimension 1 and dimension 2) showing the one-class classification of the nucleosynthesis data when the model is trained on either $r$ (top panels) or $s$ data (bottom panels) (OCC-L model here). Cases trained on $r$ include Ba, lanthanides (Z=57-70), and either (a-i) Er or (a-ii) Pb. The cases trained on $s$ also include either (b-i) Er or (b-ii) Pb. The ML model classification for the stars is also shown. For comparison, the color-coding of each star represents its JINAbase classification of which the ML model is never informed.
\label{fig:OCC_4panel}}
\end{figure*}

While in most cases stellar assignments for minimal and maximal separation  mirror one another, some assignments do change after reaching perfect separation on the training data. When this occurs the ROC curve procedure to calculate the threshold can be applied scanning from the 0 to 1 direction, as well as 1 to 0, leading to two tight threshold values near the histogram peaks which are then averaged to produce a single threshold. These tight constraints lead to 0.5 as the threshold value, which is common for binary classification.

Overall, the BC does well in recognizing $r$ and $s$ processes and agrees with the JINAbase definition 82\% of the time (see Table~\ref{tab:allresults}). However, since minimal separation is the last point where a ROC curve analysis gives a single threshold, a later epoch will require threshold averaging after perfect separation, which can move the threshold toward the simple value of 0.5. Therefore since here the threshold procedure can vary, using a BC for such a strongly separating case may prevent a clear identification of some stars. 

\section{One-class classifier}\label{sec:oneclass}

We next move towards a one class classifier (OCC) which is trained on data of only one type. It can then be used to determine whether other data are in or out of the training class. For this purpose, we use an autoencoder with a two-dimensional latent space which lets us peer into how the ML model has grouped the data before deconstructing it for the autoencoder output. 

The OCC model is driven by a Mean Square Error loss function, and, similar to BC, its architecture has an input layer of 9 nodes taking in the abundance values, but this time only from one class (only $r$ or $s$). We consider both a larger (OCC-L) and a smaller (OCC-S) model. For OCC-L(S), the input layer fully connects to the first encoder layer with 64 (32) nodes, which then connects to the second encoder layer with 32 (16) nodes, which connect to a “bottleneck” layer with 2 nodes that provide us with the latent space information. The “bottleneck” layer then feeds into decoder layers mirroring those on the encoding side, giving 9 output nodes which reconstruct the 9 input features.

We first train the OCC with the same $r$-process data considered by the BC (top panels of Fig.~\ref{fig:OCC_4panel}). 
We then use a one-class support vector machine (SVM)\footnote{\url{https://scikit-learn.org/stable/modules/outlier_detection.html}} \citep{oneclassSVM} 
with a non-linear kernel (RBF), which relies on a covariance estimation to come up with a decision boundary. We adjust SVM hyperparameters, including kernel coefficient and the upper bound on the fraction of training errors, such that the boundary encloses as many points from the training data as possible whilst still being smooth and continuous. The decision boundary, used only to interpret data previously unseen by the OCC, can be understood analogously to the threshold obtained from a ROC curve analysis in the BC case (so it is typical that not all points in the training data fall inside the boundary due to tails in the distribution). After training, latent values for data which the OCC has never seen before (e.g. training on $r$ data implies the $s$ simulations and the stellar data to be test sets) are computed. Data with values that lie within the boundary are considered to be the same class as the OCC training set and data with values outside the boundary are not within the class. 

The light blue symbols in Fig.~\ref{fig:OCC_4panel} show the latent values for $s$-process simulation data. All points lie outside the $r$-process decision boundary. Finally, we confront the trained network with the stellar abundances and evaluate whether they are considered compatible with $r$ or $s$ groupings in the latent space. We repeat this procedure by producing models that are instead trained solely on the $s$ data (bottom panels of Fig.~\ref{fig:OCC_4panel}) with $r$ being the test data.

\section{Overall classifications: \MakeLowercase{\textit{r}} vs. \MakeLowercase{\textit{s}}}\label{sec:overallclassrs}

\begin{table*}[h!]
\centering
\begin{tabular}{|p{0.5cm}|p{3.75cm}||p{0.95cm}|p{0.95cm}|p{0.95cm}|p{0.95cm}|p{0.95cm}|p{0.95cm}|p{0.95cm}|p{0.95cm}|p{0.8cm}|p{0.8cm}|}
\hline
\multicolumn{12}{|c|}{Stellar classifications with different ML methods} \\
\hline
Star \# & Name, Ref. & BC-L (Er,Pb) & BC-S (Er,Pb) & BC-L (Er,Pb) & BC-S (Er,Pb) & OCC-L (Er,Pb) & OCC-S (Er,Pb) & OCC-L (Er,Pb) & OCC-S (Er,Pb) & ML Overall & JINA-Base \\
% \hline
\cline{3-4} \cline{5-6} \cline{7-8} \cline{9-10} 
& & \multicolumn{2}{c|}{Min. separation} & \multicolumn{2}{c|}{Max. separation} & \multicolumn{2}{c|}{Trained on r} & \multicolumn{2}{c|}{Trained on s} & &\\ [-1em]
\hline 

3 & HD196944, ROE14b &  s,s & s,s & s,s & s,s & $\notr$,$\notr$ & $\notr$,$\notr$ & s,s & s,s & s & s \\
4 & SDSSJ134913.54-022942.8, BEH10 &  r,s & r,s & r,r & r,r & $\notr$,$\notr$ & $\notr$,$\notr$ & s,s & $\nots$,s & s & s \\
9 & HE0414-0343, HOL15 &  r,s & r,r & s,s & s,s & r,$\notr$ & r,$\notr$ & $\nots$,$\nots$ & $\nots$,$\nots$ & r & s \\
10 & HE0448-4806, HAN15 &  r,r & r,r & r,s & r,s & r,r & r,r & $\nots$,$\nots$ & $\nots$,$\nots$ & r & r \\
12 & HE2258-6358, PLA13 &  r,s & r,r & r,s & r,s & r,$\notr$ & r,r & $\nots$,$\nots$ & $\nots$,$\nots$ & r & s \\
16 & BD-012582, ROE14b &  r,s & r,s & s,s & s,s & $\notr$,$\notr$ & r,$\notr$ & $\nots$,s & $\nots$,s & s & s \\
22 & CS22892-052, ROE14b &  r,- & r,- & r,- & r,- & r,- & r,- & $\nots$,- & $\nots$,- & r & r \\
24 & CS22947-187, ROE14b &  r,s & r,s & r,s & r,s & $\notr$,$\notr$ & r,$\notr$ & $\nots$,$\nots$ & $\nots$,$\nots$ & r & s \\
25 & CS22881-036, ROE14b &  s,s & s,s & s,s & s,s & $\notr$,$\notr$ & $\notr$,$\notr$ & $\nots$,$\nots$ & $\nots$,$\nots$ & s & s \\
26 & CS22879-029, ROE14b &  -,s & -,s & -,s & -,s & -,$\notr$ & -,$\notr$ & -,s & -,s & s & s \\
33 & CS29513-014, ROE14b &  r,s & r,s & s,s & s,s & $\notr$,$\notr$ & $\notr$,$\notr$ & s,s & s,s & s & s \\
38 & CS22945-024, ROE14b &  r,s & r,s & r,s & r,s & $\notr$,$\notr$ & $\notr$,$\notr$ & s,s & s,s & s & s \\
40 & CS29495-042, ROE14b &  s,s & s,s & s,s & s,s & $\notr$,$\notr$ & $\notr$,$\notr$ & s,s & s,s & s & s \\
48 & SDSSJ103649.93+121219.8, BEH10 &  s,- & s,- & s,- & s,- & $\notr$,- & $\notr$,- & $\nots$,- & $\nots$,- & s & r \\
49 & LP625-44, AOK02b &  r,s & r,s & r,s & r,s & $\notr$,$\notr$ & $\notr$,$\notr$ & s,s & s,s & s & s \\
53 & CS29491-069, HAY09 &  r,- & r,- & r,- & r,- & r,- & r,- & $\nots$,- & $\nots$,- & r & r \\
57 & BD-16251, HON04 &  r,- & r,- & r,- & r,- & r,- & r,- & $\nots$,- & $\nots$,- & r & r \\
59 & HD186478, HON04 &  r,- & r,- & r,- & r,- & r,- & r,- & $\nots$,- & $\nots$,- & r & r \\
60 & HE2229-4153, SIQ14 &  r,- & r,- & r,- & r,- & r,- & r,- & $\nots$,- & $\nots$,- & r & r \\
62 & CS31062-050, JOH04 &  r,s & r,r & r,s & r,s & r,$\notr$ & r,$\notr$ & $\nots$,$\nots$ & $\nots$,$\nots$ & r & s \\
63 & CS31082-001, HIL02 &  r,r & r,r & r,r & r,r & r,r & r,r & $\nots$,$\nots$ & $\nots$,$\nots$ & r & r \\
64 & CS22896-154, CAY04 &  r,- & r,- & r,- & r,- & r,- & r,- & $\nots$,- & $\nots$,- & r & r \\
66 & BS17569-049, CAY04 &  r,- & r,- & r,- & r,- & r,- & r,- & $\nots$,- & $\nots$,- & r & r \\
70 & CS22953-003, ROE14b &  r,- & r,- & r,- & r,- & r,- & r,- & $\nots$,- & $\nots$,- & r & r \\
71 & HE2252-4225, MAS14 &  r,- & r,- & r,- & r,- & r,- & r,- & $\nots$,- & $\nots$,- & r & r \\
75 & HE0240-0807, SIQ14 &  r,- & r,- & r,- & r,- & r,- & r,- & $\nots$,- & $\nots$,- & r & r \\
78 & HD221170, IVA06 &  r,r & r,r & r,r & r,r & r,r & r,r & $\nots$,$\nots$ & $\nots$,$\nots$ & r & r \\
93 & HE1219-0312, HAY09 &  r,- & r,- & r,- & r,- & r,- & r,- & $\nots$,- & $\nots$,- & r & r \\
94 & HE0524-2055, SIQ14 &  r,- & r,- & r,- & r,- & r,- & r,- & $\nots$,- & $\nots$,- & r & r \\
95 & HD115444, HON04 &  r,- & r,- & r,- & r,- & r,- & r,- & $\nots$,- & $\nots$,- & r & r \\
96 & HD108317, ROE12b &  r,- & r,- & r,- & r,- & r,- & r,- & $\nots$,- & $\nots$,- & r & r \\
99 & BD+082856, JOH02a &  r,- & r,- & r,- & r,- & r,- & r,- & $\nots$,- & $\nots$,- & r & r \\
105& HD119516, ROE10 &  r,- & r,- & r,- & r,- & r,- & r,- & $\nots$,- & $\nots$,- & r & r \\
106 & CS30306-132, HON04 &  r,- & r,- & r,- & r,- & r,- & r,- & $\nots$,- & $\nots$,- & r & r \\
108 & CS30315-029, SIQ14 &  r,- & r,- & r,- & r,- & r,- & r,- & $\nots$,- & $\nots$,- & r & r \\
110 & BD+173248, COW02 &  r,r & r,r & r,r & r,r & r,r & r,r & $\nots$,$\nots$ & $\nots$,$\nots$ & r & r \\
112 & HE2327-5642, MAS10 &  r,- & r,- & r,- & r,- & r,- & r,- & $\nots$,- & $\nots$,- & r & r \\
113 & HD6268, ROE14b &  r,- & r,- & r,- & r,- & r,- & r,- & $\nots$,- & $\nots$,- & r & r \\

\hline
\end{tabular}
\caption{The classification found by each ML set-up explored in this work and well as the overall ML classification given the results of all approaches. The JINAbase stellar classification is also given for comparison (obtained via the $r$, $s$, and $i$ data filters on the web interface (\url{https://jinabase.pythonanywhere.com})). We use $\nots$ or $\notr$ to denote when the one-class classifier does not place the star in the $s$ or $r$ categories respectively. We use ``-" to denote when a star lacked a reported abundance for either Er or Pb. Note JINAbase still label stars without a reported Pb abundance to be $s$ based on Ba, Fe, and Eu.\label{tab:allresults}}
\end{table*}

\begin{table*}[t!]
\centering
\hskip-3cm % Added this the \centering was not centering
\begin{tabular}{|p{0.8cm}|p{5.1cm}||p{1.2cm}|p{1.2cm}|p{1.2cm}|p{1.2cm}|p{1.2cm}|p{1.2cm}|}
\hline
\multicolumn{8}{|c|}{Stellar classifications with different ML methods} \\
\hline
Star \# & Name, Ref. & OCC-L (Er,Pb) & OCC-S (Er,Pb) & OCC-L (Er,Pb) & OCC-S (Er,Pb) & ML Overall & JINA-Base \\
% \hline
\cline{3-6}
& & \multicolumn{2}{c|}{Trained on r} & \multicolumn{2}{c|}{Trained on s} & &\\ [-1em]
\hline 
20 & HE1405-0822, CUI13 & $\notr$,r & r,r & $\nots$,$\nots$ & $\nots$,$\nots$ & r & i \\
27 & SDSSJ091243.72+021623.7, BEH10 & r,$\notr$ & $\notr$,$\notr$ & s,s & $\nots$,s & s & i \\
28 & HE2148-1247, COH13 & -,$\notr$ & -,$\notr$ & -,$\nots$ & -,$\nots$ & ? & i \\
30 & HE0338-3945, JON06 & r,$\notr$ & r,$\notr$ & $\nots$,$\nots$ & $\nots$,$\nots$ & ? & i \\
41 & HE0243-3044, HAN15 & r,$\notr$ & r,$\notr$ & $\nots$,$\nots$ & $\nots$,$\nots$ & ? & i \\
\hline
\end{tabular}
\caption{Same notation as Table~\ref{tab:allresults}, now showing the five $i$-process stars (according to JINAbase) considered in this work. We use ``?" to denote when a star was never selected for a group or when its selection has effectively been canceled by a non-selection.\label{tab:iresults}}
\end{table*}

Here we assemble all of our results from considering two classification methods (binary and one-class) as well as two network sizes (small and large) for both the cases training with Er and the cases with Pb instead. For the binary cases, we report results at both minimal and maximal separation. For the one class cases, we report results from training on $r$ data alone versus training on $s$ data alone. Results are shown in Table~\ref{tab:allresults}.

We look across each row in the table to evaluate the overall ML classification for each star. When these overall classifications are compared to the JINAbase classification, we find that they agree 33/38 times ($\sim$87\% of the time). While this is encouraging, the goal of this investigation is to lay the groundwork towards using ML to classify the enrichment of stars so we are not necessarily looking to match the standard classification 100\% of the time (since it is possible for elemental ratio thresholds to 
look past other features that may be important in capturing the nature of the star). Therefore it is interesting to highlight the 5 stars that differ from the JINAbase classifications as perhaps the ML model has captured a more thorough picture of its classification. These 5 stars are: 9 = HE0414-0343, 12 = HE2258-6358, 24 = CS22947-187, 48 = SDSSJ103649.93+121219.8, and 62 = CS31062-050. Interestingly, the first observational data for star 24 from 1995 has this case formerly considered an $i$-process star by the JINAbase classification. Stars 9, 12, 24 and 62 are considered $s$ stars by JINAbase, but we predict them to be $r$ in nature, especially in the one-class cases trained on $s$ data where the ML consistently suggests these are not enriched by the $s$ process. These stars would thus be interesting for future follow-up observational campaigns.

\section{Including \MakeLowercase{\textit{i}}-process stars and nucleosynthesis calculations}\label{sec:iprocess}

\begin{figure*}[t!]
\centering
\includegraphics[scale=0.525]{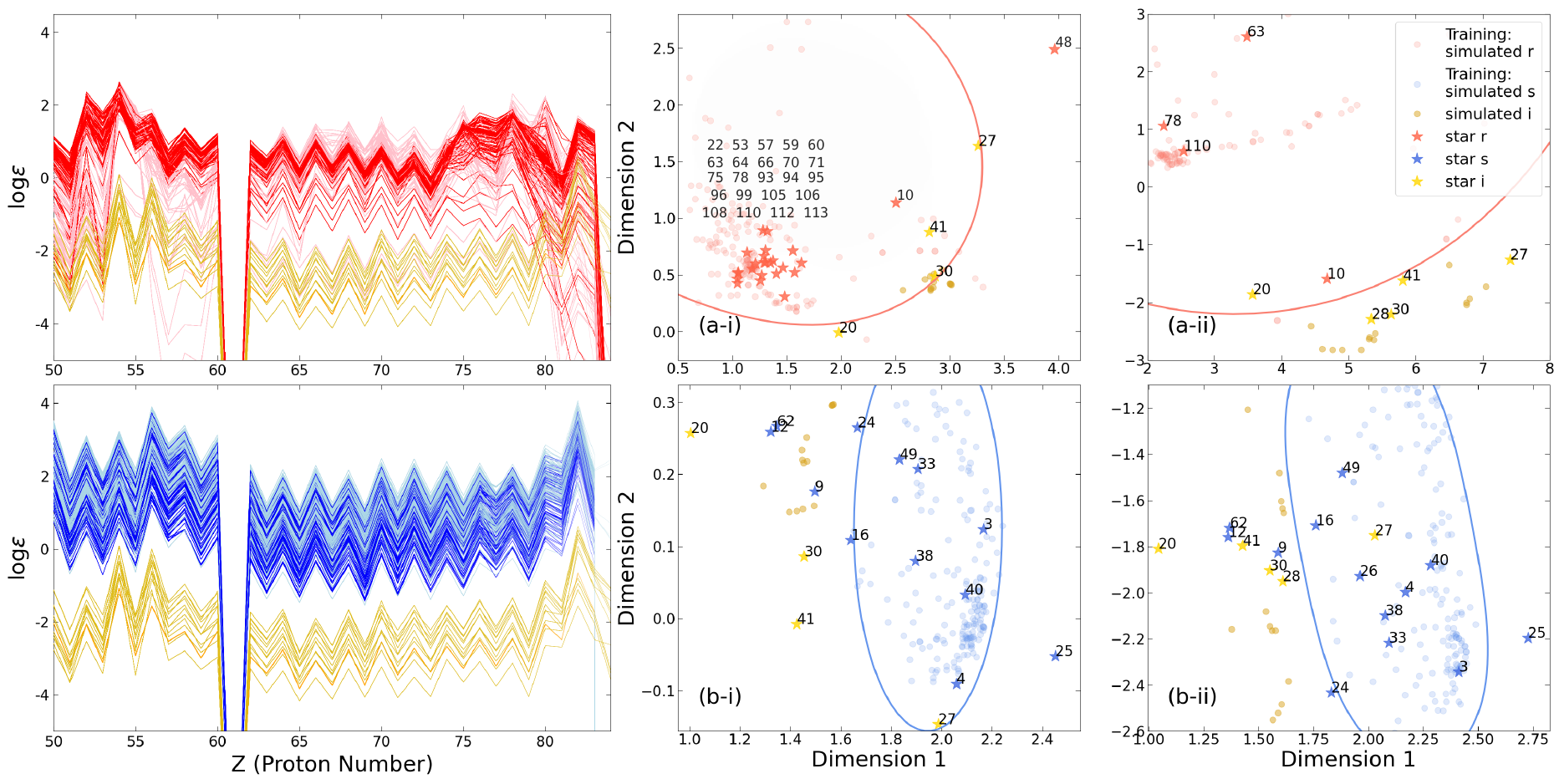}
\caption{(Left panels) Same as Fig.~\ref{fig:rsstars} but now with $i$-process nucleosynthesis calculations (yellow) shown for comparison; (Middle and right panels) Same as Fig.~\ref{fig:OCC_4panel} but with the $i$-process calculations and stars currently labeled as ``i" run through the OCC's trained on $r$ (top panels) or $s$ (bottom panels).
\label{fig:OCC_6panel_withi}}
\end{figure*}

Literature studies of the $i$ process have recently ramped up due to its ability to potentially match stellar abundances that supposedly cannot be accommodated by the $r$ process. However stellar patterns rarely include measurements of third $r$-process peak elements such as Ir (Z=77) and Pt (Z=78) (see Fig.~\ref{fig:rsstars}). Without third-peak information, some $i$-process stars with only lanthanide and lead observations could well match some $r$-process calculation ratios, and thus may actually have been enriched by the $r$ process instead. To perform a first study of $i$-process stars using our ML machinery we built for the $r$ and $s$ cases, we perform nucleosynthesis calculations with PRISM that follow the one-zone model approach of \cite{Dardelet,Tl208} in which $i$-process elements are built-up from iron and nickel when exposed to a neutron density of $10^{15}$ cm$^{-3}$. We utilize the two distinct $i$-process cases that reach the main $r$-process region reported in \cite{Benoitiprocess,Paveliprocess} (their [Fe/H] = -1.55 and [Fe/H] = -2.3 cases) as benchmarks for our PRISM calculations, and then vary our initial metallicites and the time we expose the system to neutrons to produce patterns within the range seen from this simulation group. These $i$-process calculations, including the two benchmarks calculations from \cite{Benoitiprocess,Paveliprocess}, are shown in Fig.~\ref{fig:OCC_6panel_withi} and compared with the bank of $s$ and $r$ calculations we used to train our ML models.

Table~\ref{tab:iresults} reports results for the stars that are labeled $i$ process by JINAbase (defined via 0 $<$ [La/Eu] $<$ 0.6 and [Hf/Ir]$\sim$1) which have reported values for the elements used as training features. Note that the database still label stars without a reported Ir abundance to be $i$ based on the first criterion. Table~\ref{tab:iresults} highlights that compared to the $r$ and $s$ cases in Table~\ref{tab:allresults}, the stars labeled as $i$ by JINAbase tend to have a more varied classification prediction depending on the ML set-up (e.g. Er versus Pb). 

Fig.~\ref{fig:OCC_6panel_withi} shows where $i$ data (used as test data) lie in our OCC latent space. Interestingly, the OCC latent space highlights that the ML classifier does not find the $i$-process calculations to belong in the $s$ group. Additionally, four of the five stars (20, 28, 30, and 41) are found to be strongly outside the $s$ group. The latent space suggests something different however for $i$ star 27, where almost all cases trained in $s$ data suggest this star to be in the $s$ class. Thus star 27 (SDSSJ091243.72+021623.7) could be interesting to reinvestigate with future observations as it may be connected to $s$ enrichment. In the same vein, the OCC reveals that some of the stars originally classified as $s$ process by JINAbase (stars 9, 12, and 62) align well with the region of latent space where the $i$-process calculations are found and so may in fact be of $i$-process origins. Regarding $r$ stars, star 10 (HE0448-4806) lies away from most other $r$ stars in the latent space and towards the $i$-process calculations. Thus star 10 would be of interest for future investigations of whether some $r$ stars are actually connected to $i$ enrichment (note that most, but not all, of our classifiers labeled star 10 as $r$).

 When considering the $i$ versus $r$, the distinction becomes less obvious than in the $i$ versus $s$ case.  Although most of the $i$-process calculations fall outside the $r$-process region in the latent space, they come close to the border in some cases. For three of the $i$-process stars that the ML does not associate with $s$ (stars 20, 30, and 41), at least one of the OCC set-ups finds the star within the $r$ region. Particularly for star 20 (HE1405-0822) our OCC models almost always assign this $i$ star to the $r$ class, thus suggesting a connection to $r$-process enrichment. Note none of $i$ stars considered report abundances between Z=73-81 and so the Ir, Pt ratios that may more definitively discern $r$ from $i$ are not known. Also note the very different results for how outside of the $r$ region the $i$ calculations appear when Er versus Pb is considered. This confirms Pb plays an important role in differentiating processes.
 
There is one $i$-process star considered here, star 28 = HE2148-1247, which our one-class ML never finds to be consistent with an $r$ or $s$ grouping in the latent space. These results provide another independent indication that, given currently available observational data and hydrodynamic simulations, there are stars which do not align with $r$ or $s$ process alone and an $i$ process is likely needed, as has been highlighted in recent literature. However our results also demonstrate that care should be taken in assigning a star to be of $i$-process origins since some may be more consistent with $r$-process predictions than has previously been recognized. 

\section{Conclusions}\label{sec:conclusions}

We have, for the first time, applied ML to classifying stars as having been enriched by the astrophysical neutron capture processes of $r$, $s$, or $i$. The application of ML to classify stars in this work focused on taking first steps by demonstrating the utility for $r$ vs $s$ cases. We demonstrated the promise that this method has as a new approach towards stellar classification, which is currently done via simple elemental ratio thresholds. We also demonstrated some of the challenges of this approach in that some stars were not consistently labeled to be in a given class depending on the threshold procedure and ML set-up. Interestingly, although they do not always agree on their assignments, overall the OCC and BC perform similarly in terms of their overall match with JINAbase classifications. However, the OCC latent space provides additional valuable insights regarding how strongly a given abundance pattern (from either stellar and simulation data) correlates with a class.

This work highlights that the $r$, $s$, and $i$ processes need to be carefully disentangled given that many stars currently classified in one group were favored by the ML to be in another. Interestingly some the most hotly contested cases were those currently labeled as $i$-process stars. Overall our ML models find a few $i$ stars to fall within the latent space of the $r$ class and one of them to fall within the $s$ class. However, given current abundance data, one of the $i$-process stars considered (star 28 = HE2148-1247) consistently resists joining either of the two groups and identified more with the latent space region in which the $i$-process calculations were found. Future work in which the ML is actually trained on $i$-process calculations by using a larger set from current leading simulation groups would be needed to make more solid claims about $i$ stars. 

Here we trained the ML models on the abundances of Ba, the lanthanides, and Pb. We found that training with Pb at times gave a different prediction than in the case with just Ba and lanthanides. However it is the $r$-process third peak at Z=77,78 that may be more of the smoking gun between the $s$/$i$ and $r$ patterns, and only a few stars have reported observations for these elements. Thus efforts by the observational community to produce more complete abundance patterns would be informative to future ML classification studies. 

Future applications of ML to stellar classification are many, including studies which take into account variations in nuclear physics inputs for the $r$-process calculations to evaluate if some nuclear models are more capable than others of aligning with stellar abundances in ML determined latent spaces. Further, these methods could be applied to differentiate between astrophysical sites of a given process (e.g. merger dynamical ejecta, accretion disk wind ejecta, MHD supernovae ejecta, collapsar ejecta). As hydrodynamic simulations evolve, ML approaches can then adapt their classification predictions accordingly. For instance, MHD supernovae simulations presently predict the production of mostly elements lighter than the lanthanides considered here \citep{Reichert}, but if that picture were to change ML approaches could easily incorporate this new training data. Thus this work introduces the idea of turning ML loose on the stars to recognize abundance patterns and lays the early foundations of utilizing ML to classify stellar enrichment.

%\begin{acknowledgments}
%N.V., Y.W., M.L., and K.M. acknowledge the support of the Natural Sciences and Engineering Research Council of Canada (NSERC). TRIUMF receives federal funding via a contribution agreement with the National Research Council (NRC) of Canada. M.P.K. was supported by the National Science Foundation under Grant No. PHY-2012865.
%\end{acknowledgments}

\section*{acknowledgments}
N.V., Y.W., M.L., and K.M. acknowledge the support of the Natural Sciences and Engineering Research Council of Canada (NSERC). TRIUMF receives federal funding via a contribution agreement with the National Research Council (NRC) of Canada. M.P.K. was supported by the National Science Foundation under Grant No. PHY-2012865.

\section*{Appendix \\ Machine learning methods}

We utilize the PyTorch library\footnote{\url{https://link.springer.com/chapter/10.1007/978-3-030-57077-4_10}}, an open-source ML library for Python programs, to build artificial neural networks (ANN or NN) and train the classifiers in both the BC and OCC methods. Our networks are comprised of multiple layers of nodes, each of which perform two functions - collecting inputs (usually from the previous layer) and generating outputs (usually to be fed into the following layer). Before providing the nucleosynthesis data to the NN, we normalize data of the nine features from each simulation by scaling to its respective Eu abundance (scaling each so that log$\epsilon$(Eu)=0). For cross verification, we also tested with other commonly-used normalization methods with scikit-learn\footnote{\url{https://dl.acm.org/doi/abs/10.1145/2786984.2786995}}, including StandardScaler and MinMaxScaler from the sklearn.preprocessing package. The classifiers, when trained on data normalized with these different normalization methods, shows reasonable agreement. In this work we report the results from the Eu-normalization method. 

We train the classifier with a mostly balanced dataset with approximately equal entries of simulated $r$- and $s$-process data. The availability of $s$-process data, here 187 patterns from the grid simulations released by two independent nucleosynthesis groups, was the limiting factor in the size of our datasets. In order to balance the amount of $s$ vs $r$ data considered, we mapped the set of 420 Radice et al. patterns down to 124 before combining them with the 64 cases from Just el al. This was achieved by running all 420 cases through the STUMPY fast pattern matcher which finds the nearest neighbor of each pattern. The set of patterns selected as neighbors are then considered to be sufficient to capture the full set (here mapping 420 cases to 302). This process is repeated until it “saturates” since at some point the neighbors pair up and keep selecting one another (e.g. here 302 maps to 261, 261 to 250, and 250 to 248 where it saturates). At this point the pairs permit half the data to be safely captured by its partner and so the set can be cut in half. The resulting set of 124 excludes many redundant patterns from the original set thus maximizing the diversity of patterns the ML model is exposed to. After training with the reduced 124 Radice + 64 Just set, we checked how the trained models classify the omitted 296 cases from Radice et al. They were found to be consistent with the $r$-process histograms (BC) and decision boundaries (OCC).

These nucleosynthesis calculation datasets then provide our ML model with two key pieces of information - features and targets. The features are the 9 elemental abundances considered. Targets are labels, here either ``r" or ``s", to indicate whether the data is from an $r$- or $s$-process simulation. The target labels are then binary encoded into 0 and 1, which corresponds to ``r" and ``s" respectively. All entries, each containing its nine feature values and one target label, are then shuffled and randomly split, where 60\% are used as the training set, 20\% the validation set, and 20\% the test set. The dataset used for one-class classification only consists of entries of the same class, either ``r" or ``s", on which the network is exclusively trained. Again, we take 60\%, 20\%, and 20\% of the randomly shuffled data for training, validation, and test, respectively. 

\subsection*{Binary classification methods}

For the BC results presented in this work, we used a shallow neural network architecture (one input layer, one hidden layer, and one output layer). We also tested other deeper NNs by increasing the number of hidden layers and varying the numbers of nodes in each layer; this gave results that are generally consistent with those from the shallow NNs. 

\begin{figure}
    \centering
    \includegraphics[width=1.08\linewidth]{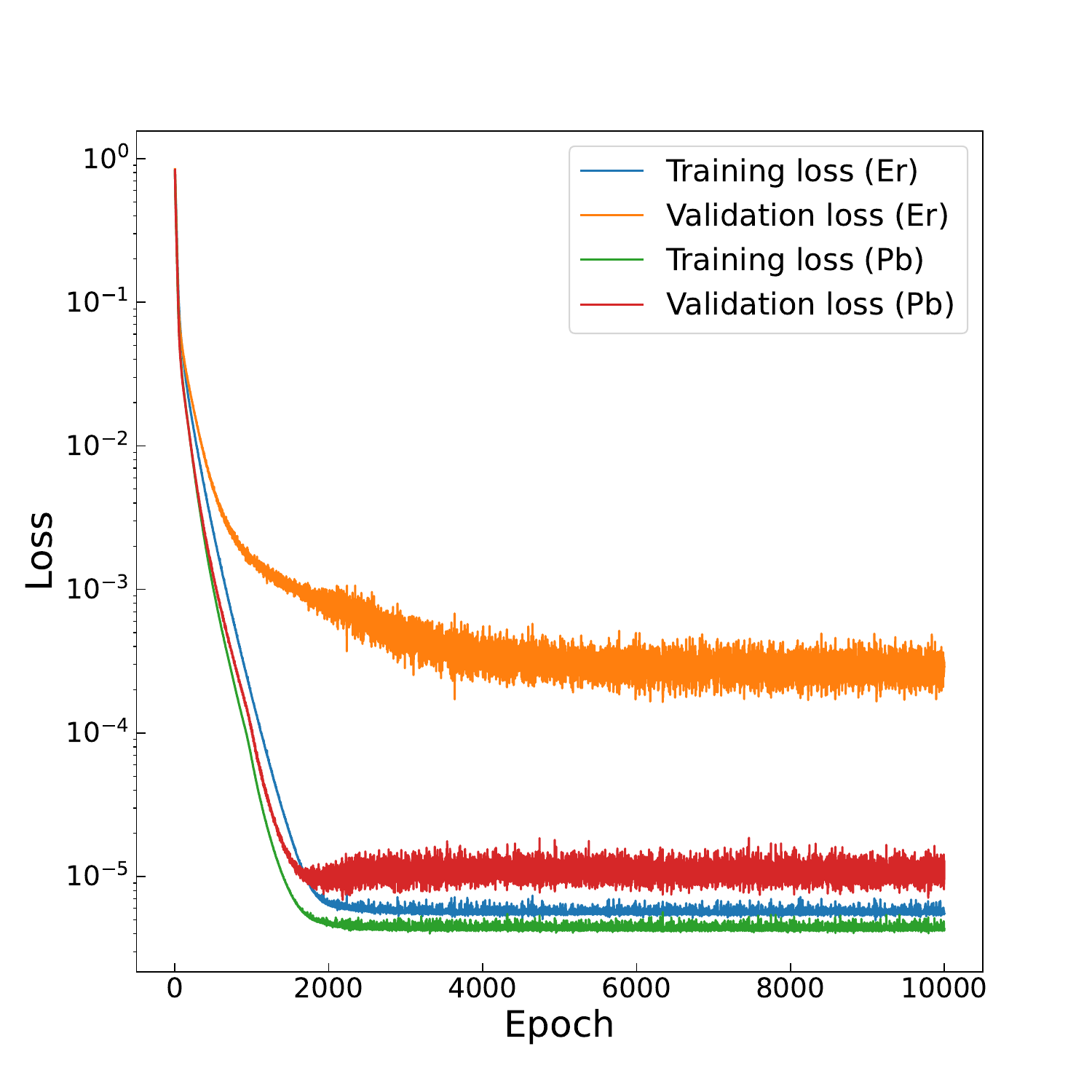}
    \caption{Training and validation losses as the BC-L network evolves over 10,000 epochs, using 9 elements (features) up to Er and up to Pb respectively.}
    \label{fig:losscurves}
\end{figure}

The NN is trained using the training portion of the dataset. Some weights are assigned to the connections between nodes, and an initial loss is calculated with a Binary Cross Entropy loss function (torch.nn.BCE). As the network is trained over iterations of epochs, the weights on the connections are adjusted with the aim to reduce the training loss with the Adam optimizer (torch.optim.Adam)\footnote{\url{https://arxiv.org/abs/1412.6980}}. The connection weights, at every epoch, are also applied using the validation set as the input, and the validation loss is calculated and tracked. The validation loss, observed in comparison with the training loss, serves as an indicator of when the model has reached its best trained stage, which occurs when the validation loss is no longer being reduced as the training continues, even if the training loss continues to improve. Here we iterate through 10,000 epochs and define the model to be at its best-trained state when the validation loss (Binary Cross Entropy) is at its global minimum. Two examples of training and validation losses are included in Figure \ref{fig:losscurves}, where one training-validation loss pair indicates overtraining and the other pair does not. 

The classification results are reported in Table \ref{tab:allresults}. We refer to the best-trained state of the model as the state of maximal separation. We also examine the specific epoch after which the histograms of the predicted values of the two classes from the training set no longer overlap, which we refer to as the minimally separated state. Minimal separation can be understood in terms of optimizing the true positive rate (TPR) and false positive rate (FPR), where

\begin{equation}
    TPR = \frac{TP}{\text{Actual Positive}} = \frac{TP}{TP+FN},
\end{equation}
\begin{equation}
    FPR = \frac{FP}{\text{Actual Negative}} = \frac{TN}{TN+FP}.
\end{equation}

\begin{figure}
\centering
\includegraphics[width=0.55\linewidth]{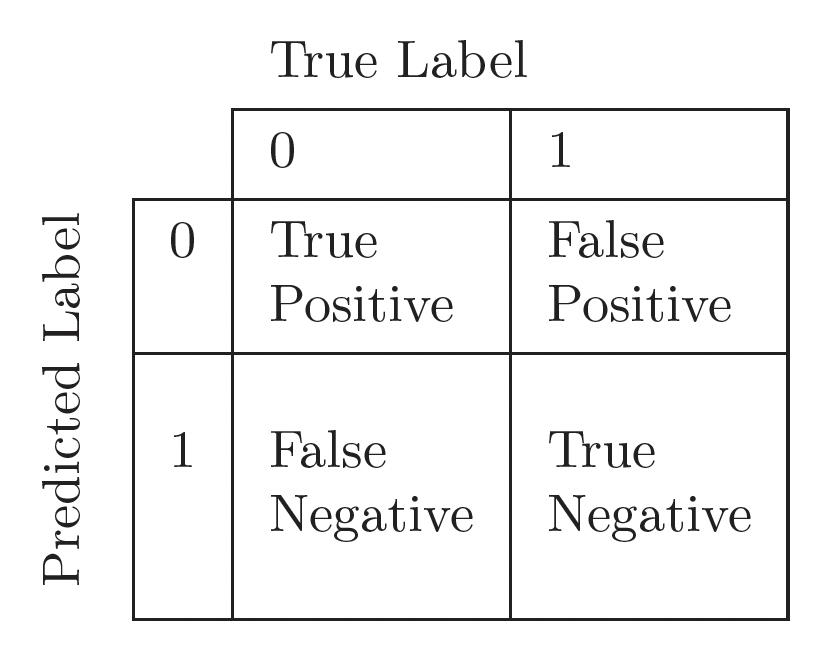}
\caption{Confusion Matrix, where true positive, false positive, false negative, and true negative are defined based on their true and predicted labels. \label{fig:cm}}
\end{figure}

\begin{figure}[htb!]
    \centering
    \includegraphics[width=1\linewidth]{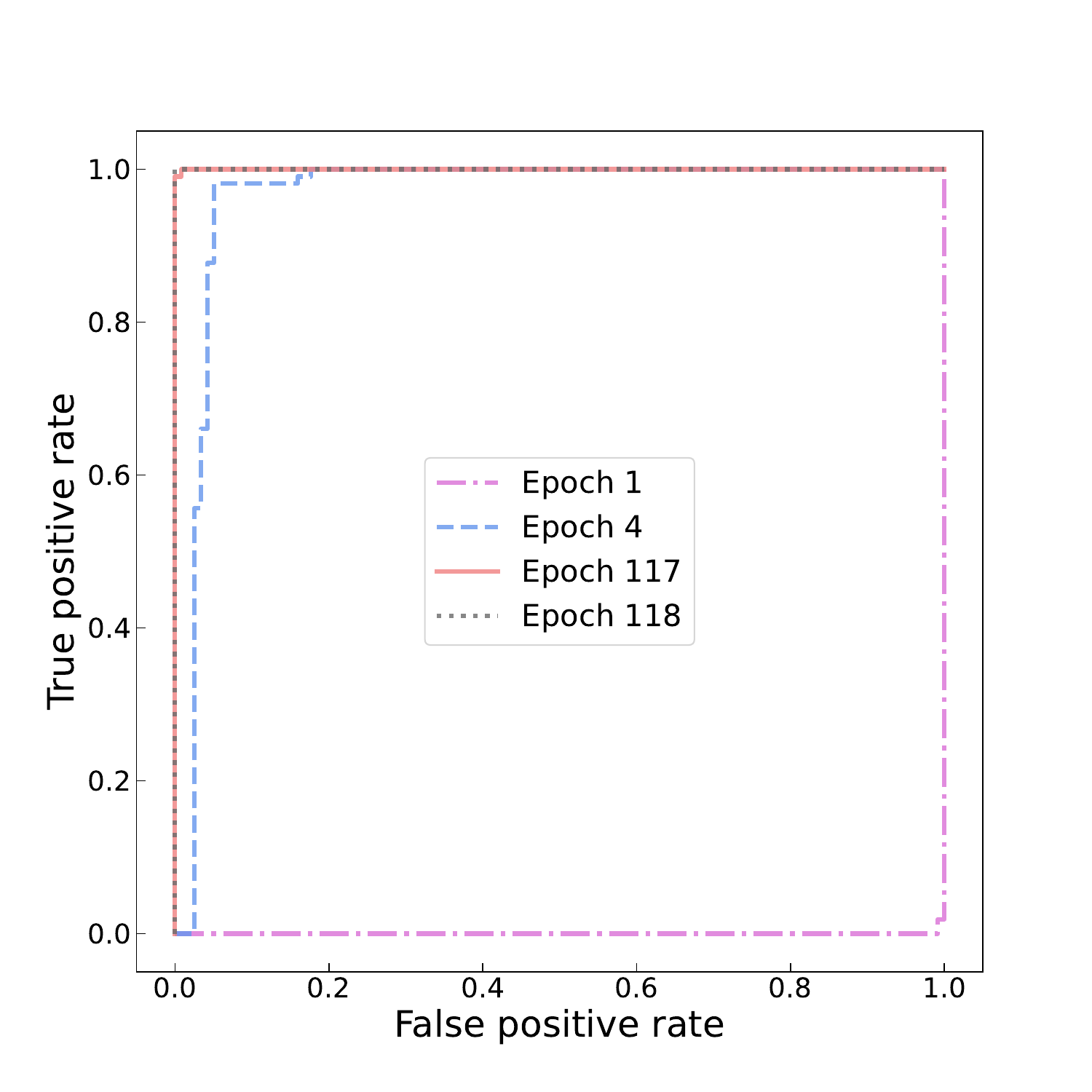}
    \caption{ROC curves using the predicted outputs using the BC-L network trained on 9 elements up to Er shown at various epochs, where epoch 118 corresponds to the state of minimum separation.}
    \label{fig:ROC}
\end{figure}

\noindent At each epoch, we span a range of possible thresholds between 0 and 1. At each threshold, the cases with outputs smaller than the threshold value are assigned the predicted label 0 and the cases that are larger are assigned 1. The definitions of true positive (TP), false positive (FP), true negative (TN), and false negative (FN) are summarized by the confusion matrix shown in Figure \ref{fig:cm}. We find the optimal threshold value at which TPR is maximized and FPR is minimized. One commonly-used technique to assess the optimal threshold is by plotting the receiver operating characteristic (ROC) curve, comprising points with TPR, FPR evaluated at all the thresholds sampled. We find the point on the ROC curve with the shortest distance to the upper left corner where TPR = 1 and FPR = 0, and take the threshold corresponding to that point as the optimal threshold. The area under the curve (AUC) is a metric that indicates the classifier's ability to distinguish between the two classes, where AUC = 1 indicates complete separation between the two classes. Evaluating the ROC curve for outputs from each epoch, we can interpret the epoch $n-1$ as the final instance at which the AUC of the ROC is smaller than 1, and take epoch $n$ as the point of minimal separation. It should be noted for the ROC curves evaluated at epochs prior to complete separation (i.e. epochs 1 to $n-1$), there is one unique point with minimal distance to the (0, 1) point and therefore one unique optimal threshold. However, for epoch $n$ onward, multiple sampled thresholds yield TPR = 1 and FPR = 0, leading to multiple points on the ROC curve having the same minimal distance of 0 to the upper left corner and thereby a range of threshold values that are equally optimal. Under such circumstances, we take the minimal and maximal values in the series of eligible thresholds that yield TPR = 1 and FPR = 0, and take the average between them as the optimal threshold reported, yielding the thresholds reported for both minimum and maximal separation in Figure \ref{fig:BC4panel}. Figure \ref{fig:ROC} shows the ROC curves at a few specific epochs using the BC-L network training on 9 elements up to Er, where epoch 118 is the point of minimal separation. The AUC for epoch 118 (and onwards) equals 1 and indicates complete separation.

\subsection*{One-class classification methods}

\begin{figure}[t!]
    \centering
    \includegraphics[width=1\linewidth]{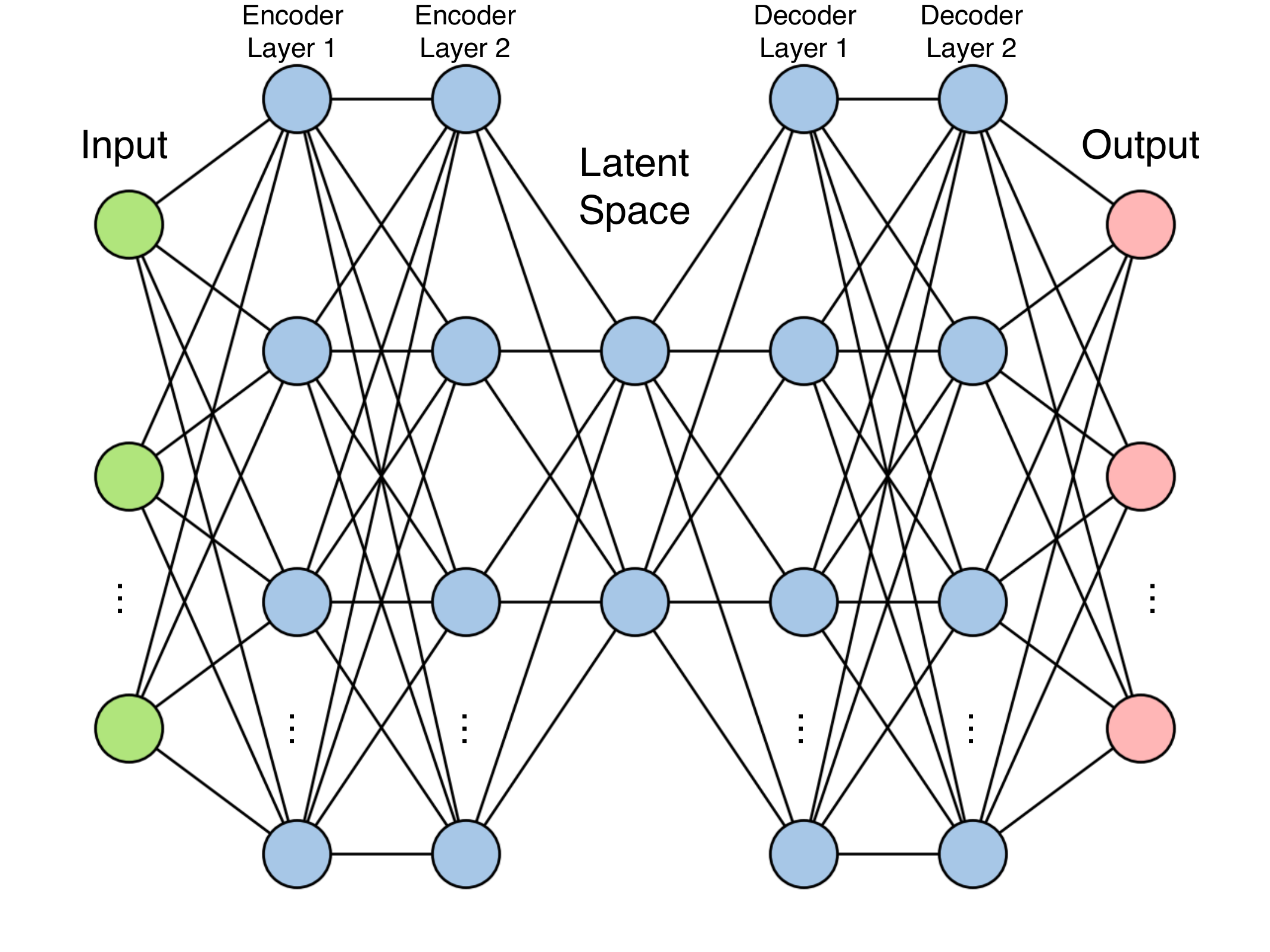}
    \caption{Schematic of the autoencoder architecture used for OCC.}
    \label{fig:AE}
\end{figure}

For one-class classifier (see e.g. \cite{oneclass}), we use auto-encoder models to first encode the nine input values into a latent space representation through a series of encoder layers, then decode them back to nine output values through decoder layers. The encoder part of the network, in essence, acts as a dimensionality compressor, and the decoder portion re-expand the information from the latent space into the original dimensions. The encoder and decoder structure mirror each other, and the input and output layers have the same number of nodes corresponding to the number of features the network is trained on. The latent space layer often is of a lower dimension, hence it is also commonly referred to as the bottleneck layer. Figure \ref{fig:AE} illustrates the structure of an autoencoder containing a two-dimensional latent space, as is the case in this work. 

For the OCC we use a Mean Square Error loss function (torch.nn.MSELoss) and the Adam optimizer. We train the model on a dataset containing one class only, referred to as the training class. We end all runs after 30,000 epochs as we see clear signs of overtraining based on the trend in the validation loss in all cases. We again take the best-trained state as the state of the model at which the validation loss is at its global minimum. 

We use the weights from the best-trained state to obtain the latent space values for data in the training class. We then take a dataset on which the model is not informed, referred to as the testing class, and apply the same weights to obtain the latent space values. For example, when training the classifier on simulated $r$-process data, the training class is the ``r" class on which the model has been informed, and the testing class includes simulated $s$-process data as well as stellar data, all of which the trained model has been blind to. We then obtain a boundary around the latent space values of the training class using a one-class Support Vector Machine (SVM). Since the SVM boundary is not informed on the latent space values of the testing class, it serves as a means to evaluate any testing class on which the model is not informed, such as the stellar data, without ever having to train on such cases.

\bibliography{MLrefs.bib}
\bibliographystyle{aasjournal}

\end{document}